\documentclass{moriond}
\usepackage{graphicx}
\usepackage{amsmath}
\usepackage{subcaption}

\newcommand{\tb}{{\bar{t}}}
\newcommand{\qb}{{\bar{q}}}

\def\be{\begin{equation}}
\def\ee{\end{equation}}
\def\bea{\begin{eqnarray}}
\def\eea{\end{eqnarray}}

\newcommand{\Photo}{}

\begin{document}
\vspace*{4cm}
\title{Progress Towards Two-loop QCD Corrections to $pp\to t\bar{t} j$}

\author{Simon Badger}

\address{Dipartimento di Fisica and Arnold-Regge Center, Università di Torino, and INFN, Sezione di Torino, Via P.\ Giuria 1, I-10125 Torino, Italy}

\maketitle\abstracts{
We present the status of on-going efforts to compute the two-loop virtual corrections to $pp\to t\bar{t} j$ in the leading colour
approximation. We review the recent study of the master integrals and their differential equations and present some ideas on modern techniques that can be
applied to provide complete evaluation of helicity amplitudes containing information on top quark decays in the narrow width approximation.
}

\section{Introduction}

Top quark pair production in association with an additional jet is a high
priority process with current and future LHC experiments.  This signal forms
a significant fraction, around 50\%, of all top-quark pair events and has been
studied extensively by both ATLAS and CMS.~\cite{CMS:2016oae,ATLAS:2018acq,CMS:2020grm,CMS:2024ybg} The normalized distributions for $t\bar{t}j$ have been shown to be highly
sensitive to the top-quark mass~\cite{Alioli:2013mxa} and therefore precision
predictions are in high demand.

While next-to-leading order (NLO) QCD corrections have been known for many
years,~\cite{Dittmaier:2007wz} the next-to-next-to-leading order (NNLO) QCD
corrections are currently unavailable due to the missing two-loop virtual
contributions. The computational complexity of these amplitudes presents a
formidable challenge to current theoretical methods and new techniques are
required. New methods using finite field arithmetic,~\cite{vonManteuffel:2014ixa,Peraro:2016wsq} highly optimized
integration-by-parts reduction~\cite{Gluza:2010ws,Ita:2015tya,Larsen:2015ped,Wu:2023upw} and improved analytic understanding of
multi-scale Feynman integrals~\cite{Gehrmann:2015bfy,Gehrmann:2018yef,Chicherin:2020oor,Chicherin:2021dyp,Abreu:2023rco} have made considerable impact on $2\to3$
scattering processes. Recent highlights have been the completion of full colour corrections to $pp\to 3j$ and $pp\to\gamma+2j$ at NNLO.~\cite{Badger:2023mgf,Agarwal:2023suw,DeLaurentis:2023nss,DeLaurentis:2023izi} The first steps have been taken to develop these ideas
for applications to $pp\to t\tb j$ amplitudes~\cite{Badger:2022mrb,Badger:2022hno} where promising results have
been obtained. Here we report on the status of the leading colour two-loop
amplitudes where differential equations for the master integrals have recently
been derived.~\cite{Badger:2024fgb}

\section{Amplitudes in the leading colour limit}

Since the amplitudes are of an extremely high complexity, we start by
considering the limit of a large number of colour charges $N_c\to\infty$. There
is evidence to suggest that such an approximation to the two-loop virtual
correction would be sufficient for next-to-next-to-leading order since the size
with respect to double-real and real-virtual contributions is often small using
standard infra-red subtraction schemes.

Helicity amplitude techniques have been extremely successful for predictions
with massless particles. For massive particles helicity is not well defined
since one can always boost to a frame in which helicity is flipped. We can
therefore only define helicity with respect to a reference direction. Helicity
amplitude techniques can be extremely useful in the case of top-quark pair
production since it is simple to include decays in the narrow width
approximation.~\cite{Melnikov:2010iu}

The leading colour decomposition of the loop amplitudes, $\mathcal{A}^{(L)}$ can be written as
\begin{align}
  \mathcal{A}^{(L)}&(1_\tb, 2_t, 3_g, 4_g, 5_g) = g_s^{3+2L} N_c^L \nonumber\\
      &\sum_{\sigma\in S_3} (t^{a_{\sigma(3)}} t^{a_{\sigma(4)}} t^{a_{\sigma(5)}} )_{i_2}^{\bar{i}_1}
                            A^{(L)}_1(1_\tb, 2_t, \sigma(3)_g, \sigma(4)_g, \sigma(5)_g)
  \label{eq:colourdecomp_2t3g}
\end{align}
in the gluon channel and
\begin{align}
  \mathcal{A}^{(L)}&(1_\tb, 2_t, 3_q, 4_\qb, 5_g) = g_s^{3+2L} N_c^L \bigg\{ \nonumber\\
	     &\delta^{\bar{i}_4}_{i_1}(t^{a_{5}})^{\bar{i}_2}_{i_3} A^{(L)}_{1}(1_\tb, 2_t, 3_\qb, 4_q, 5_g)
      + \delta^{\bar{i}_3}_{i_2}(t^{a_{5}})^{\bar{i}_4}_{i_1} A^{(L)}_{2}(1_\tb, 2_t, 3_\qb, 4_q, 5_g) \bigg\}
  \label{eq:colourdecomp_2t2q1g}
\end{align}
in the (light) quark channel. Here we have used $g_s$ as the strong coupling, $N_c$ as the number of colours and $t^a$ are the $SU(N_c)$ generators.
Due to the high gluon luminosity at the LHC, the gluon channel will dominate the contributions. At leading colour the external
particles always appear with a cyclic ordering and so the top-quark pairs
remain adjacent. This has the effect of minimising the number of massive
internal propagators and therefore leads to considerably simpler Feynman
integrals. Feynman integrals with internal masses can give rise to elliptic
integrals from two-loops onwards. These structures are currently of great
interest to mathematicians and there is an active field of research aiming to
obtain an understanding suitable for collider physics applications.
Unfortunately, for complicated kinematics such as $pp\to t\bar{t}j$, the
technology for numerical evaluation of such functions is not available at the
same level as it is for more standard situations that can be written in terms of
multiple polylogarithms.

\begin{figure}[t!]
	\centering
	\begin{subfigure}{0.25\linewidth}
		\includegraphics[width=\linewidth]{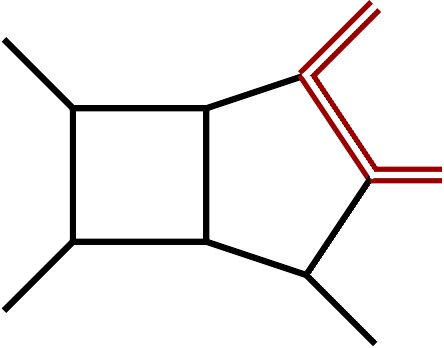}
		\caption{Topology ${\rm PB}_A$.}
		\label{fig:PBttjA}
	\end{subfigure}
	\begin{subfigure}{0.25\linewidth}
		\includegraphics[width=\linewidth]{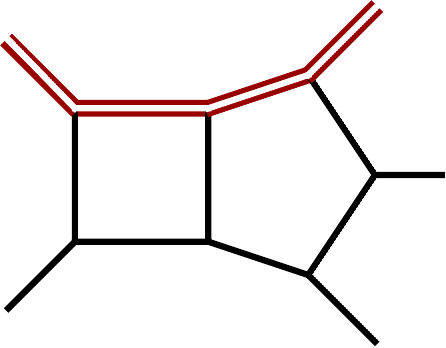}
		\caption{Topology ${\rm PB}_B$.}
		\label{fig:PBttjB}
	\end{subfigure}
	\begin{subfigure}{0.25\linewidth}
		\includegraphics[width=\linewidth]{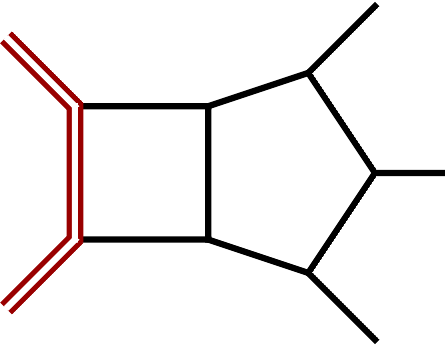}
		\caption{Topology ${\rm PB}_C$.}
		\label{fig:PBttjC}
	\end{subfigure}\\
	\begin{subfigure}{0.25\linewidth}
		\includegraphics[width=\linewidth]{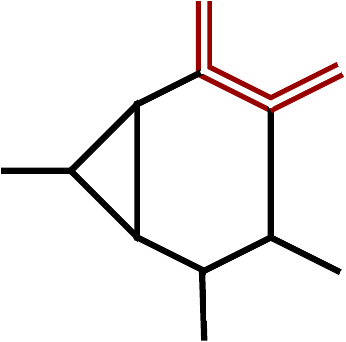}
		\caption{Topology ${\rm HT}_A$.}
		\label{fig:HTttjA}
	\end{subfigure}
	\begin{subfigure}{0.25\linewidth}
		\includegraphics[width=\linewidth]{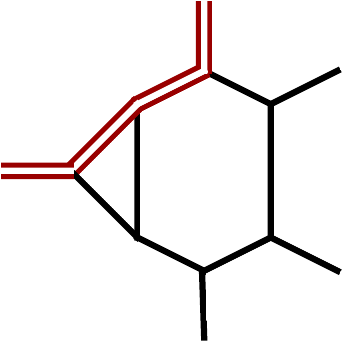}
		\caption{Topology ${\rm HT}_B$.}
		\label{fig:HTttjB}
	\end{subfigure}
	\begin{subfigure}{0.25\linewidth}
		\includegraphics[width=\linewidth]{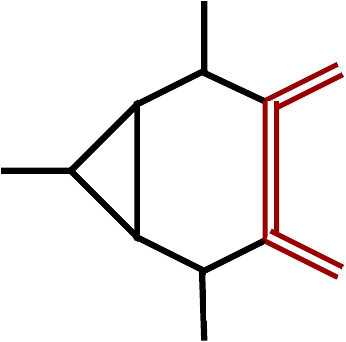}
		\caption{Topology ${\rm HT}_C$.}
		\label{fig:HTttjC}
	\end{subfigure}
  \caption{The six integral families appearing in the tensor integral
  representation of $pp \rightarrow t\bar{t}j$ at two-loops in the leading colour limit.
  Black lines denote massless particles and red double-lines denote massive
  particles. The master integrals of all three hexagon-triangle topologies can be mapped to the pentagon-box topologies.}
	\label{fig:ttj_allfams}
\end{figure}

From analysis of the Feynman diagrams of the two-loop leading colour amplitudes, one
can find nine different integral families. Three topologies are formed from bubble insertions
into one-loop pentagon graphs and can be trivially mapped into the three
`hexagon-triangle' topologies. After solving integration-by-parts identities to
determine the master integrals, one can determine that there are no master
integrals in the top sector of the hexagon-triangles and so these master
integrals can also be mapped into the `pentagon-box' families. The families are
shown in Figure \ref{fig:ttj_allfams}.

\section{Evaluation of the pentagon-box master integrals}

In the recent preprint article~\cite{Badger:2024fgb}, differential equations
(DEs) for all pentagon-box master integrals have been presented. Bases of
master integrals were found for topologies ${\rm PB}_A$ and ${\rm PB}_C$ where
the DEs can be written in terms of logarithmic one-forms. Contrary to all other
pentagon-box configurations considered so far, topology ${\rm PB}_B$ was shown
to have two problematic sectors. One could only be rotated in to
$\epsilon$-factorized form by introducing a nested square-root while another was
shown to contain elliptic structures. As a result the strategy of expanding the
integrals into a basis of stable pentagon functions which can be evaluated
efficiently is not straightforward. As an alternative, we demonstrated that
numerical values could be obtained in a reasonable time from generalized series
expansions of the DEs.~\cite{Francesco:2019yqt,Hidding:2020ytt} Topology ${\rm PB}_B$ was not obtained in
$\epsilon$-factorized form but as a quadratic polynomial in $\epsilon$ which could be
represented compactly in terms of linearly independent one-forms.

We find that topologies ${\rm PB}_A$ and ${\rm PB}_C$ have 71 and 79
independent one-forms respectively, all of which can be written as $\mathrm{d}
\log$ forms. Topology ${\rm PB}_B$ has 135 independent one-forms of which 72
can be written as $\mathrm{d} \log$ forms. Evaluation times per segment of the
path between points in the generalized series expansion are around 5 times
longer for ${\rm PB}_B$ than for ${\rm PB}_A$ and ${\rm PB}_C$ but, at around 1
minute per point, feasible for phenomenological applications. Considerations
such as numerical stability and efficient algorithms for spanning the
phasespace while minimizing the number of segments remain for further study. For
the two topologies admitting $\mathrm{d} \log$ representations, it is
straightfoward to derive analytic representations in terms of multiple
polylogarithms or pentagon functions that would provide fast and stable numerics.

\section{Outlook}

Having established a set of master integrals suitable for the complete
amplitude in the planar limit that can be reliably evaluated numerically, the
next step is to obtain their rational coefficients for each amplitude. This
could be attempted though analytic reconstruction from finite field evaluations (modulo a prime number), for example using the \texttt{FiniteFlow} package,~\cite{Peraro:2019svx} as has been successful for other $2\to3$ scattering
processes. Analytic forms for rational coefficients may well be extremely
complicated and it may be useful to also consider numerical approaches. While
this step will require the reduction of high rank tensor integrals, the finite
field strategy combined with optimized IBP systems does appear to be viable.
The unknown aspect of the evaluations is the average time per phase-space point
which will certainly be considerably more expensive than massless $2\to3$
processes. There are many unexplored optimizations for efficient evaluation of
the integrals and numerical interpolation techniques that lead us to believe
the two-loop virtual corrections can be obtained via this strategy in the not
too distant future.

\section*{Acknowledgments}

I would like to thank my collaborators Matteo Becchetti, Colomba Branccacio, Nicol\`o Giraudo, Heribertus Bayu Hartanto and Simone Zoia. This project has received funding from the European Union's Horizon Europe research and innovation programme under the Marie Skłodowska-Curie grant agreement No.~101105486, and ERC Starting Grant No.~101040760 \emph{FFHiggsTop}. This work has received funding from the Italian Ministry of Universities and Research through FARE grant R207777C4R. This research was supported in part by the Swiss National Science Foundation (SNF) under contract 200021\_212729. SB has been partially supported by the Italian Ministry of Universities and Research (MUR) through grant PRIN 2022BCXSW9.

\end{document}